# Photostimulated phosphor based image plate detection system for HRVUV beamline at Indus-1 synchrotron radiation source

K. Haris, Param Jeet Singh[a], Aparna Shastri[a]*, Sunanda K.[a], Babita K.[a], S.V.N. Bhaskara Rao[a], Shabbir Ahmad and A. Tauheed

*Department of Physics, Aligarh Muslim University, Aligarh 202002, India*

[a]*Atomic & Molecular Physics Division, Bhabha Atomic Research Centre, Mumbai 400085, India*

**ABSTRACT**

A high resolution vacuum ultraviolet (HRVUV) beamline based on a 6.65 meter off-plane Eagle spectrometer is in operation at the Indus-1 synchrotron radiation source, RRCAT, Indore, India. To facilitate position sensitive detection and fast spectral recording, a new BaFBr:Eu$^{2+}$ phosphor based image plate (IP) detection system interchangeable with the existing photomultiplier (PMT) scanning system has been installed on this beamline. VUV photoabsorption studies on Xe, $O_2$, $N_2O$ and $SO_2$ are carried out to evaluate the performance of the IP detection system. An FWHM of ~ 0.5 Å is achieved for the Xe atomic line at 1469.6 Å. Reproducibility of spectra is found to be within the experimental resolution. Compared to the PMT scanning system, the IP shows several advantages in terms of sensitivity, recording time and S/N ratio, which are highlighted in the paper. This is the first report of incorporation of an IP detection system in a VUV beamline using synchrotron radiation. Commissioning of the new detection system is expected to greatly enhance the utilization of the HRVUV beamline as a number of spectroscopic experiments which require fast recording times combined with a good signal to noise ratio are now feasible.

*Author for correspondence: Email: ashastri@barc.gov.in; Tel: +91-22-25590343; Fax: +91-22-25502652

# 1. Introduction

Spectroscopy in the vacuum ultraviolet (VUV) region has always posed a challenge due to scarcity of suitable sources and detection methods. With the advent of dedicated synchrotron radiation (SR) sources, highly intense and tunable beams of VUV photons are now available which have facilitated a variety of experiments which were either impossible or very difficult to perform with traditional laboratory sources [1-3]. Detection of VUV radiation in the early years was carried out using specially prepared gelatin-free photographic glass plates like Kodak make short wavelength radiation (SWR) plates [3, 4]. Due to some inherent disadvantages like fragility, aging problems, cumbersome dark room procedures for loading and developing, etc. these plates were gradually phased out and replaced by electronic detectors like photomultiplier tubes (PMTs), microchannel plates (MCPs) and charge coupled devices (CCDs). While VUV PMTs with $MgF_2$ windows are available for direct detection of radiation down to 1100 Å, for lower wavelengths, one has to use a scintillator. Most commonly used is a sodium salicylate film which absorbs in the VUV and emits fluorescence in the visible region, which is then detected by a visible PMT [3]. In cases where position sensitive detection or imaging is desirable, MCPs and CCDs have been used. However, besides being much more expensive, these have limited wavelength coverage and resolution as compared to photographic plates. PMTs, MCPs and CCDs are also prone to electrical noise. Typically, in high resolution spectrometers, photographic plates have been replaced with PMTs using a precision scanning mechanism along the Rowland circle to cover the wavelength region of interest. This means that for achieving high resolution and good signal to noise ratios small step sizes and long integration times have to be used. Due to these reasons, SWR plates are still in use for a number of spectroscopic applications. However, since the manufacturing of these plates has been recently discontinued, an urgent necessity is created to find alternative position sensitive detectors.

The photostimulated phosphor based image plate (IP) detector is one such alternative which is inexpensive and offers many advantages over both PMTs and traditional photographic plates. This detector consists of a film of BaFBr with trace amounts of $Eu^{2+}$ coated on a flexible plastic substrate and works on the principle of photostimulated luminescence (PSL) [5]. Although phosphor based IPs had been in wide use in the X-ray region for crystallography and medical diagnostics for quite some time, their usefulness for detection in the VUV region was first demonstrated by Reader *et al*. [4] who used special autoradiographic plates mounted in a 10.7 m grazing incidence spectrometer to record spectra in the 50–550 Å region from a low inductance vacuum spark source. At about the same time, Ben-Kish *et al*. [6] also demonstrated the use of similar IPs for extreme UV (XUV) spectroscopy by recording spectra from a fast capillary discharge Z-pinch source using a 2 m Schwob-Fraenkel spectrometer in the 5–1700 Å region. IPs used by them were modified to have thinner phosphor layers than standard X-ray image plates and no protective plastic coating in order to render them useful in the VUV region. These experiments concluded that the PSL based IP has sensitivity comparable with that of photographic plates, along with a better linear response over a wider dynamic range. Subsequently, Nave *et al*. [7] showed that the usefulness of these plates can be extended to the longer wavelength side (UV region) up to 2300 Å, with some sensitivity retained even up to 3000 Å. It may be noted that all the papers cited above are on VUV emission spectroscopy. Despite the capability and advantages of IPs as VUV detectors, we do not find much literature reporting their use in synchrotron radiation source based VUV photoabsorption studies. Here, a mention must be made to the paper by M. Katto *et al*. [8] where they have compared the photoluminescence intensity from $BaFBr:Eu^{2+}$ films irradiated by synchrotron radiation and pulsed VUV lasers.

In this paper we report the implementation of an IP based detection system on the high resolution VUV (HRVUV) beamline [9] at the Indus-1 synchrotron radiation source (SRS), Raja Ramanna Centre for Advanced Technology (RRCAT), Indore, India and its application to photoabsorption studies using SR. We have augmented the 6.65 meter spectrometer to enable use of an IP based detection system interchangeably with the existing PMT based detection system. The IP is mounted on the focal plane of the spectrometer and the performance of the new system with respect to the PMT based technique is compared, especially with respect to recording time, sensitivity, reproducibility, dynamic range, resolution, etc.. Details of the design of the new system and test experiments performed are discussed in what follows.

**2. Experimental Setup**

**2.1 Brief description of beamline and PMT scanning mechanism**

The Indus-1 SRS is a 450 MeV electron storage ring with a peak photon flux of $7.2 \times 10^{11}$ photons/sec/mrad/0.1% bandwidth [10]. The HRVUV beamline at Indus-1 which has been in operation since 2011 is designed for photoabsorption studies of atoms and molecules using SR in the 1150–3500 Å wavelength range [9]. A schematic layout of the HRVUV beamline is shown in Figure 1. Details of the design and development of this beamline have been reported earlier [9]. Briefly, this indigenously developed beamline utilizes a 6.65 meter off-plane Eagle mount normal incidence spectrometer for wavelength dispersion. The spectrometer houses a single optical component, *viz.* an Al+MgF$_2$ coated concave grating with groove density 1200 l/mm, blazed at λ=1500Å giving a first order reciprocal linear dispersion of 1.24 Å/mm in the wavelength range of 1150–3500Å. With this grating mounted in the spetometer, one can record a scanning range of 160 mm corresponding to ~ 200 Å for a given central wavelength setting. A 4800 l/mm gold coated grating which covers a wavelength range of ~ 50 Å for each central wavelength setting with reciprocal linear dispersion of ~ 0.3 Å/mm is also available for higher resolution measurements. The wavelength range of interest is set by rotating the grating about the horizontal axis of the spectrometer to select a particular central wavelength ($\lambda_0$) and then translating it along this axis to focus the dispersed light onto the exit slit [11]. The angle of rotation $\alpha_0$ and translational position $X$ are given by the relations $\alpha_0 = \sin^{-1}(\lambda_0/2d)$ and $X = R(1-\cos\alpha_0)$ where d is the groove spacing and R is the radius of curvature of the grating. The entire spectrometer chamber is maintained at a vacuum of $1 \times 10^{-6}$ mbar obtained using a roots blower and two 500 l/s turomolecular pumping stations.

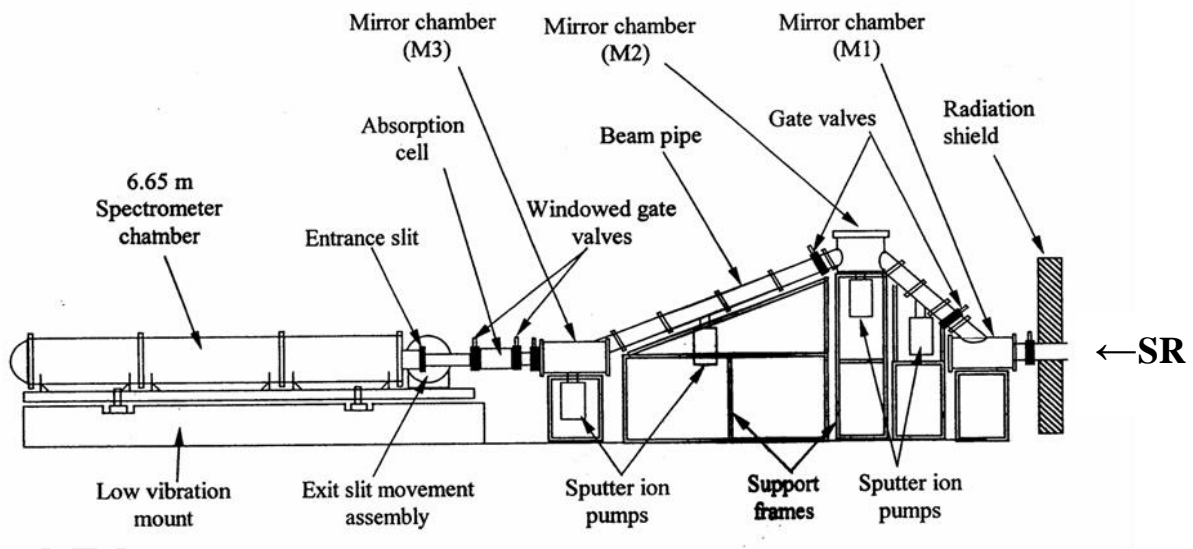

**Figure 1. Schematic layout of the HRVUV beamline at Indus-1**

For photoabsorption studies, a 0.5 m stainless steel absorption cell (*cf.* Figure 2) is placed between the entrance slit of the spectrometer and beamline fore-optics. The broad band SR beam from the Indus-1 source passes through a three cylindrical mirror fore-optics system and subsequently through the absorption cell before getting focused onto the entrance slit of the spectrometer. The cell is isolated from the spectrometer and the beamline optics by two ultrahigh vacuum (UHV) gate valves with lithium fluoride (LiF) windows mounted in them and is equipped with a gas filling system consisting of several ports for pumping, introduction of gases and pressure measurement. A system of Swagelok needle valves is used to regulate the flow of gaseous samples and capacitance manometers are used to accurately measure the sample pressure.

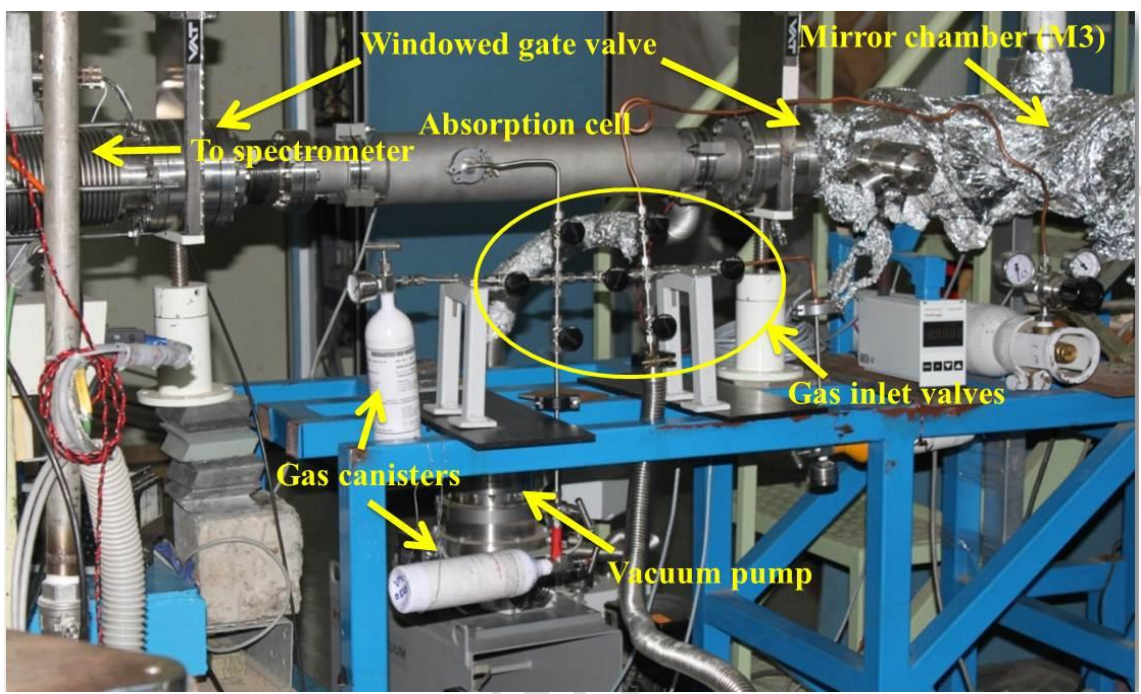

**Figure 2. Gas phase absorption cell with sample introduction system**

The dispersed radiation is detected by a sodium salicylate coated quartz window coupled with a visible PMT, which is scanned along the Rowland plane in steps of 10 μm size, with accuracy of ± 1 μm for recording the spectra [9] . The linear and rotational motion of the grating and the linear motion of PMT are controlled by a 3-axis stepper motor controller and driver system. The motor control and data acquisition system are interfaced to a PC using a VISUAL BASIC software [9]. Using this beamline, photoabsorption studies of some stable molecules have been successfully carried out [12]. However, for extending the scope of experiments that can be carried out, the existing detection system has several limitations. For instance, the lifetime of the storage ring which is ~1 hour at 100 mA injection current poses a serious problem, as the beam decay is quite drastic over the period of one scan. Moreover, at relatively low beam currents (10–50 mA), the signal to noise (S/N) ratio of the spectra recorded by the PMT detection system is rather poor, thus necessitating usage of relatively high beam currents ( > 50 mA) in order to obtain sufficiently high S/N ratios. The resultant high incident flux often causes photodissociation of the molecule being studied, and the absorption spectra may show peaks belonging to one or more dissociation products.

One of the proposed research problems on this beamline is the study of VUV absorption spectra of transient species (radicals) produced in a discharge, which requires short recording times. With the existing PMT detection system, the typical time taken to record a medium resolution (~ 0.5 Å) spectrum spanning a total spectral range of 200 Å is about 45 – 60 min (at 50 μm step size). In order to record the next scan, the PMT has to be first brought back to the home position, which takes ~ 15 min. Therefore the effective time for a single scan is at least one hour. For high S/N ratios, longer integration time is required, whereas for higher resolution, the PMT has to be scanned in smaller steps. In such cases the time required for a single scan may be as much as 3 hours. During this time, drifts in experimental conditions and SR beam decay cause the recorded spectra to be prone to errors. These factors prompted us to explore the use of an IP as an alternative detection method which is highly sensitive, can detect low signals and record maximum possible wavelength region (~ 200 Å) in a very short time (~ 1–60 s).

## 2.2 Description of the image plate detection system

Image plates or phosphor screens consist of photostimulable crystals $BaFBr:Eu^{2+}$ (phosphor) coated onto a flexible polyester substrate with a typical grain size of ~ 5 μm [4, 6]. The mechanism of the image storing process in phosphor screens has been well documented earlier [5]. The main process involved is the conversion of $Eu^{2+}$ ions to $Eu^{3+}$ on irradiation with VUV light (>6.9 eV) and formation of F centers by the excited electrons. Additionally, in lower energy regions (~ 4–6.9 eV) other mechanisms may also contribute [5]. These processes effectively form latent images which can be stored for several days until exposed to strong white light. In order to read out the stored information in IPs, they are scanned in a specially designed laser scanner. Laser excitation at 633 nm leads to photoluminescence at ~ 390 nm which is then detected by a PMT. The PMT signal measured at various points on the film produces a two dimensional map which can be readout on a computer as a digital image. The data stored in the IPs can be erased by exposing them to strong white light for about 2 minutes which makes the IP reusable several times (up to 1000 times). Several different types of IPs are available depending on the method of preparation of the phosphor film, its thickness, etc., resulting in different response/sensitivity. We have used tritium sensitive (TR) screens, designed for medical applications that use tritium for labeling. TR films are manufactured with highest grade $BaFBr: Eu^{2+}$ crystals and are not over-coated with plastic films. They are thus ideal for detection of less penetrating radiation like VUV and also offer a good combination of sensitivity and resolution. Since TR films are uncoated they require very careful handling and are prone to moisture accumulation and contamination with use. In this regard, unlike the standard X-ray IPs, in order to re-use these films a large number of times, certain precautions have to be observed, viz. careful handling to avoid scratches/fingerprints and storing in a dry and dark environment. For scanning the image, the IPs are wrapped around a drum and held in place by end clips. The drum is then scanned by rotation at a constant speed (360 rpm) which ensures linearity of the position measurements. Multiple scanning is generally used to retrieve the stored information from the IPs efficiently. Since data accumulated in each revolution are added up, the signal to noise ratio and dynamic range are enhanced. The storage phosphor scanning system (Model no. C431200) and the IPs used in the present study are procured from M/s. PerkinElmer. The OptiQuant software which is provided along with the scanner controls the scanning and is also useful for image acquisition, analysis, display and archiving.

## 2.3 Design of plate holder & modifications carried out for implementation of IP detection system

A photograph of the PMT scanning chamber in which the IP is to be placed is shown in Figure 3a. For this purpose, a plate holder was designed and fabricated with a radius of curvature of 6.65 meter so that IP lies along the focal plane of spectrograph at exit channel. The aluminum plate holder used for the preliminary studies is capable of holding a single IP of size 25 mm x 250 mm (*cf*. Figure 3b). Two thin metallic strips with screws are provided to hold the IP in place. For these experiments, the IP detector mounted on the plate holder is inserted into the chamber using a 63 CF port on top of the existing scanning chamber. For experiments which require PMT detection, it is easily possible to revert back to the scanning mechanism. Prior to taking exposures, the detection chamber is evacuated to $10^{-5}$ mbar using a turbo molecular pumping station. Vacuum isolation between the spectrometer chamber and detection (scanning) chamber is provided by a manually operated rectangular gate valve.

The disadvantages of this single IP holder are that one can mount only one film at a time. In order to take the next exposure, the detection chamber has to be vented to atmosphere, the erased IP reloaded and the chamber re-evacuated to $10^{-5}$ mbar. To overcome these limitations, an improved plate holder was designed which can hold six IPs strips at a time. The modified plate holder is essentially a hexagonal stainless steel rod with six faces of dimension 1.5 cm x 25 cm (*cf*. Fig.3c) mounted on a flange and inserted from a port on top of the scanning chamber. Each face has two railings running along the length of the rod with screws which are tightened to keep the IP strip in place. One can rotate the hexagonal rod in vacuum by means of a simple Wilson seal rotary feedthrough mechanism, such that one face at a time is exposed to the dispersed radiation from the grating. Markings on an external handle indicate which face is being exposed. Between two adjacent faces, aluminum strips have been provided to ensure that during a particular exposure, other IP's are not exposed to the dispersed radiation. Figure 3d shows the placement of the six IP strips on the drum prior to scanning. In this manner, one can record six exposures before the chamber has to be vented, thus bringing down the time required for 6 exposures from ~ 6 hours (including the pump down and venting times) to ~ 1 hour. This enables us to record spectra at different wavelength settings with the same sample pressure, thus covering a large spectral region (~ 250 Å for a given wavelength setting) within a short time (~30 min). Similarly for a given central wavelength setting, one can record spectra with varying sample conditions.

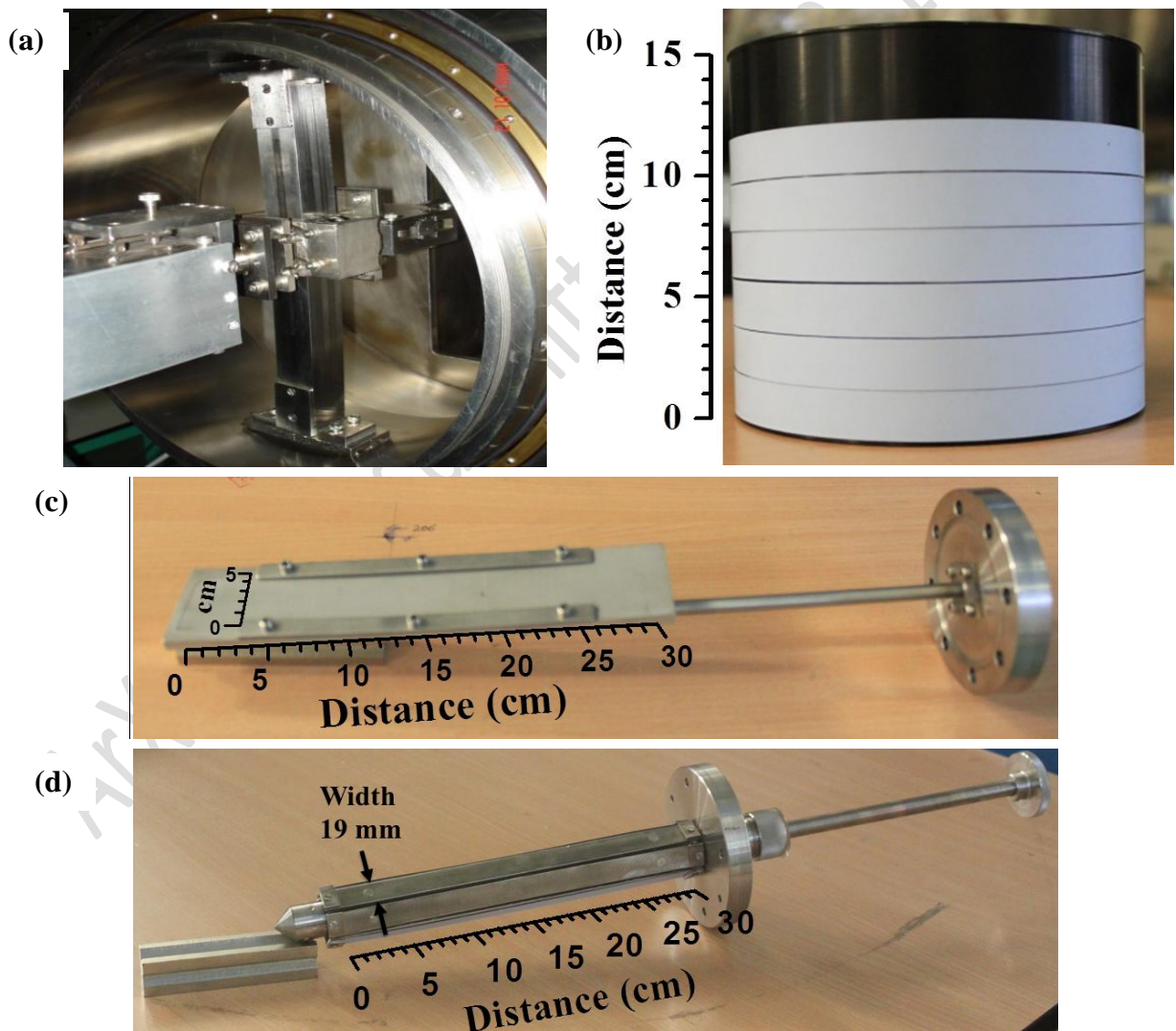

**Figure 3. (a) Scanning chamber of the 6.65 meter spectrometer (b) Scanning drum with six IP strips mounted (c) Single IP holder (d) Hexagonal multiple IP holder. Length scale is indicated in cm.**

## 3. Performance of the IP system for VUV photoabsorption studies

To estimate the performance of the IP detection system with respect to sensitivity, resolution, S/N ratio, reproducibility, etc., VUV photoabsorption spectra of a few atomic and molecular species were recorded. For these experiments, the absorption cell is evacuated to a base pressure of ~1.5x$10^{-6}$ mbar or better using a turbomolecular pumping station. The IPs are loaded on to the holder and the detector chamber is evacuated to a pressure of $10^{-5}$ mbar. After introducing the sample into the absorption cell at the desired pressure, the rectangular gate valve isolating the IP chamber from the rest of the spectrometer is opened. The duration of exposure of the IP to the SR is controlled using a pneumatic gate valve at the front end of the beamline. Exposure time (~ 5 sec to 1 min) is chosen based on the beam current and sample conditions. In case of the multi-IP holder, six such exposures can be taken under varying experimental conditions. After the exposures, the detection chamber is isolated from the spectrometer and vented to atmosphere through a special vent valve equipped with a dust filter. The IPs are mounted on the drum and scanned at 600 dpi (dots per inch) resolution, corresponding to a pixel size of 42 micron and saved as a TIF (tagged image format) file. The image is then imported into a plotting software for spectral digitization. Standard image processing routines enable conversion of the TIF image into a plot of transmitted intensity versus position (mm) which is related to wavelength through the reciprocal linear dispersion (1.24 Å/mm).

### 3.1 Xenon

Xenon is a favored atomic standard for VUV absorption studies as it has several atomic absorption lines in the region 1150–1500 Å [13]. In the present study, xenon gas of stated purity 99.99% (procured from M/s. Six Sigma Gases) was filled into the absorption cell at pressures of 0.1–0.6 mbar and exposed to synchrotron radiation for a few seconds, ~ 5–30 seconds in accordance with the beam current.

A central wavelength setting ($\lambda_c$) of 1235 Å was chosen in order to record the four Xe atomic lines at 1170.4, 1192.0, 1250.2 and 1295.6 Å. All four lines could be recorded with good intensity at a pressure of 0.6 mbar (*cf*. Figure 4a). Optimization of the spectrometer resolution was carried out using the xenon atomic line at 1469.6 Å, by varying parameters such as the focusing distance *X*, slit width, slit tilt and sample pressure. A full width at half maximum (FWHM) of 0.2 Å at 40 μm slit width is achieved using the PMT detection system while with the IP, the FWHM obtained is 0.48 Å at 30 μm slit width (*cf*. Figure 4b).

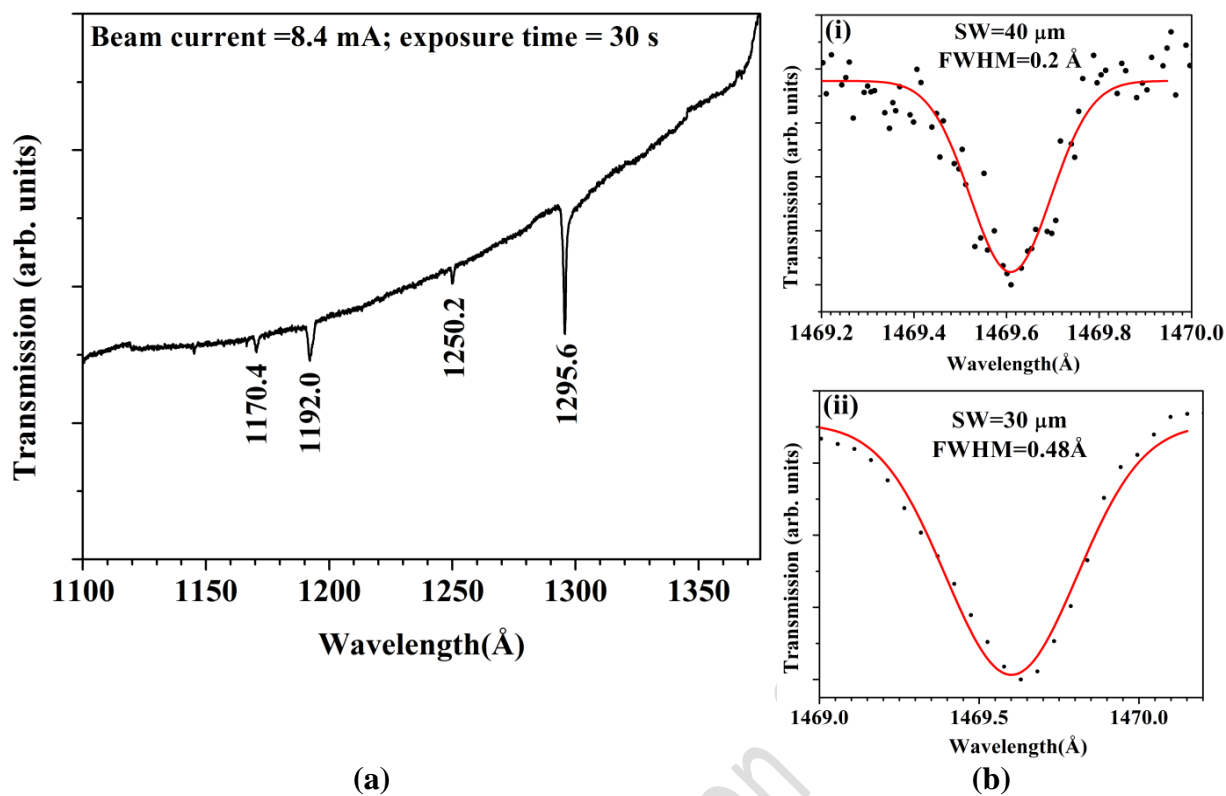

**Figure 4.** (a) Atomic absorption lines of Xe (0.6 mbar pressure; $\lambda_c$=1235 Å) in the wavelength region 1100–1370 Å recorded using IP detection system. (b) Line width of 1469.6 Å Xe line (0.1 mbar pressure; $\lambda_c$=1430 Å) using (i) PMT and (ii) IP

### 3.2 Oxygen

The VUV photoabsorption spectrum of the oxygen molecule was recorded in wavelength range of 1700–1900 Å at various pressures (1 mbar–10 mbar) and beam currents (5–80 mA). The well known Schumann Rungé bands of $O_2$ recorded at a pressure of 3 mbar using the IP with exposure time of 30 seconds at a beam current of 8 mA are shown in Figure 5a as obtained from the OptiQuant scanner. Figure 5b shows the same spectrum with the X-axis converted into wavelength scale, along with that obtained using PMT detection under similar experimental conditions. Band head assignments (v′, v″) marked on the figure (where v′ and v″ are the vibrational quantum numbers of the excited and ground states respectively) match well with earlier literature [14].

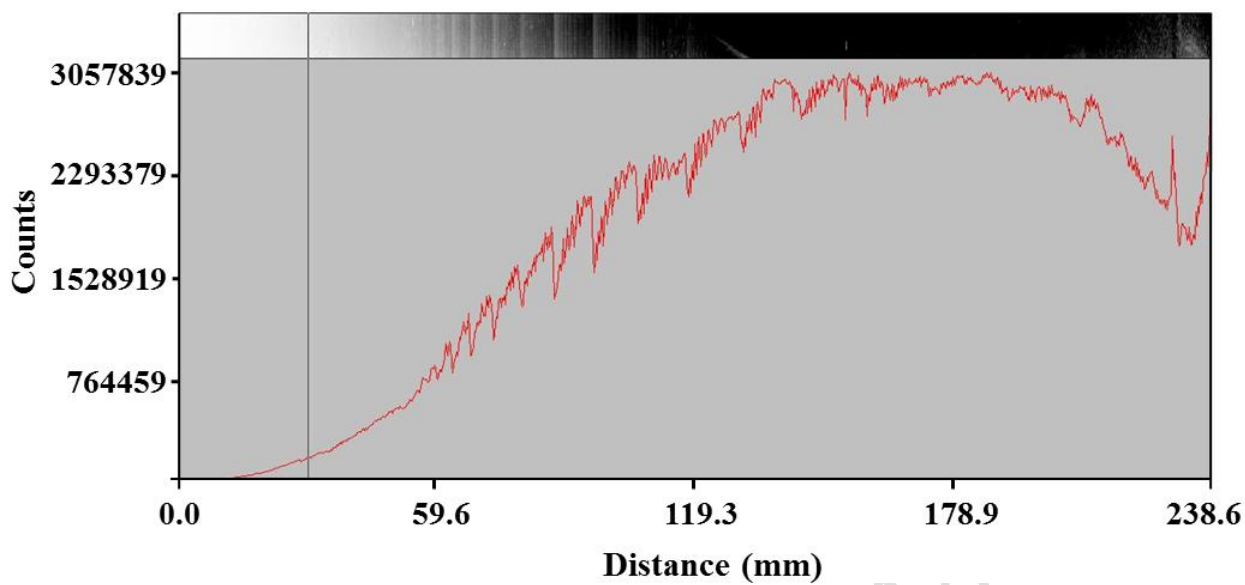

**Figure 5a.** Schumann Rungé bands of $O_2$ ( P~4 mbar) in the wavelength range 1750–1900 Å recorded on HRVUV beamline using IP, as obtained by the OptiQuant software

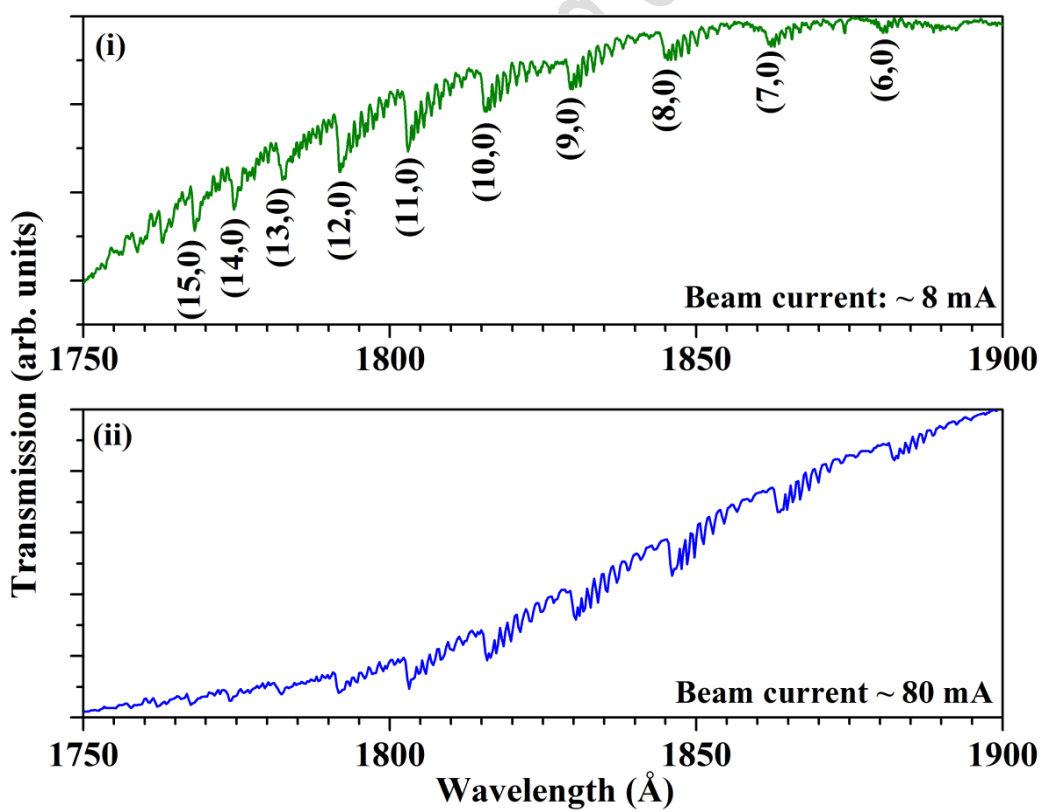

**Figure 5b.** Schumann Rungé bands of $O_2$ (P ~ 4 mbar) in the wavelength range 1750–1900 Å recorded on HRVUV beamline (i) using IP, converted to transmission versus wavelength scale and (ii) using PMT detection.

### 3.3 Sulphur dioxide

The B–X and C–X systems of $SO_2$ lie in the 1700–2300 Å region and show rich vibronic structure. The photoabsorption spectrum of $SO_2$ (0.5 mbar) in the region 1900–2150 Å recorded using the IP detection system at a beam current of 25 mA is shown in Figure 6. The spectrum in this region consists mainly of bands belonging to the C–X system. A comparison of the spectrum reported in literature [15] (dotted line in Figure 6) with the IP spectrum shows good agreement.

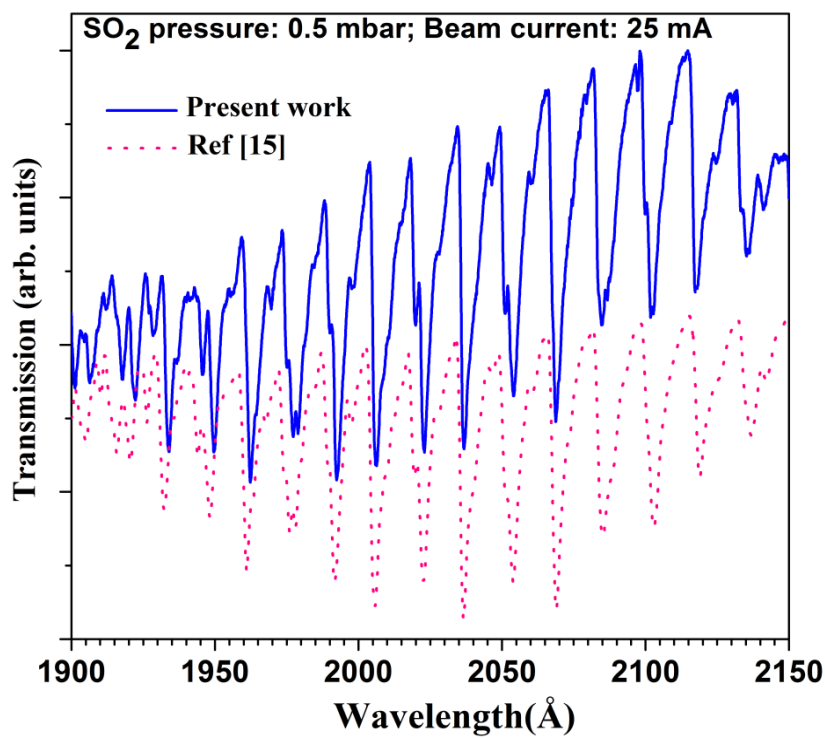

**Figure 6. Photoabsorption spectrum of $SO_2$ in the wavelength region 1900–2150 Å recorded on HRVUV beamline using IP**

### 3.4 Nitrous oxide

A pressure dependent photoabsorption study of the $C^1\Pi - X^1\Sigma^+$ system of $N_2O$ [16] recorded using the multi-plate holder is shown in Figure 7. Here, xenon was mixed with the $N_2O$ sample to serve as a calibration reference. Keeping the central wavelength fixed at 1469 Å, the sample pressure was varied from 0.01 to 5 mbar. It is observed that at pressures above 1.5 mbar, the spectrum shows saturation, while for pressures below 0.05 mbar, absorption features are very weak. The optimum pressure at which all bands in this spectral region are observed clearly is ~ 0.1 mbar. The reproducibility of spectral features is found to be within the experimental resolution. This study also demonstrates the utility of the multi-plate holder in performing multiple experiments in a short time during which parameters like beam current do not change appreciably.

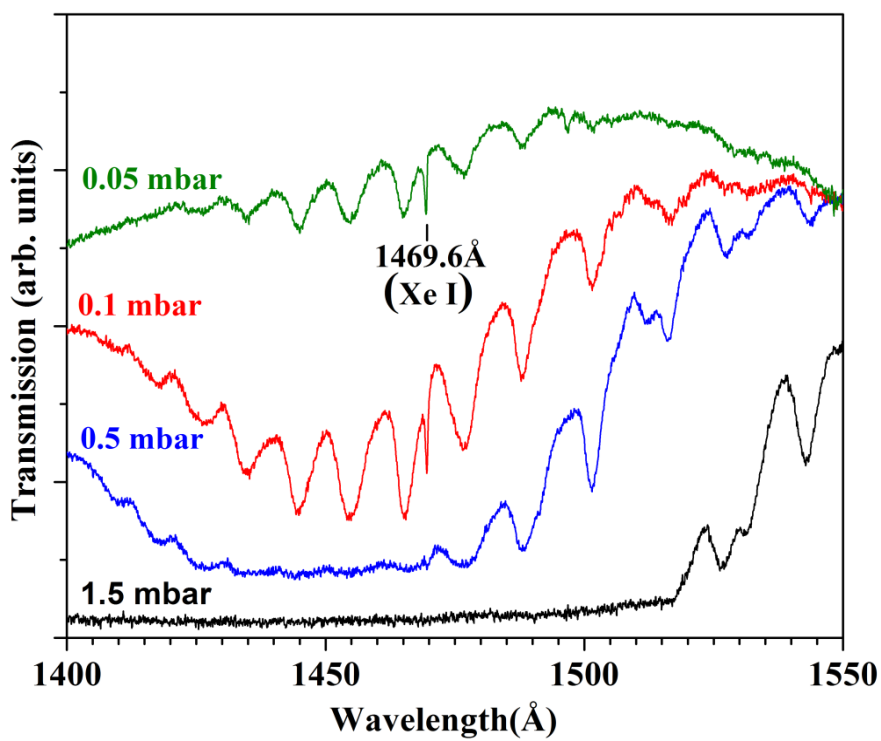

**Figure 7.** Photoabsorption spectrum of $N_2O$ in the 1400–1550 Å region at several pressures recorded on HRVUV beamline using IP

### 3.5 Emission spectrum from Cu arc

The sensitivity of IPs in the UV region ($\lambda > 2200$ Å) has been discussed in an earlier report [7]. In the present study, the arc spectrum of Cu in the 2200 – 3000 Å wavelength range was recorded using a medium quartz spectrograph at Aligarh Muslim University. Figure 8 shows the Cu arc spectrum taken at an exposure time of 15 sec. From Figure 8, it can seen that the sensitivity of the IP is quite good up to ~ 2600 Å (as evident from the high intensity of the 2618 Å line) while at higher wavelengths, there is a drastic fall in sensitivity. Nevertheless, the two prominent Cu I lines at 2766 and 2824 Å (*cf*. inset in Figure 8) are detected with weak intensity, confirming that IPs can be used effectively up to 3000 Å.

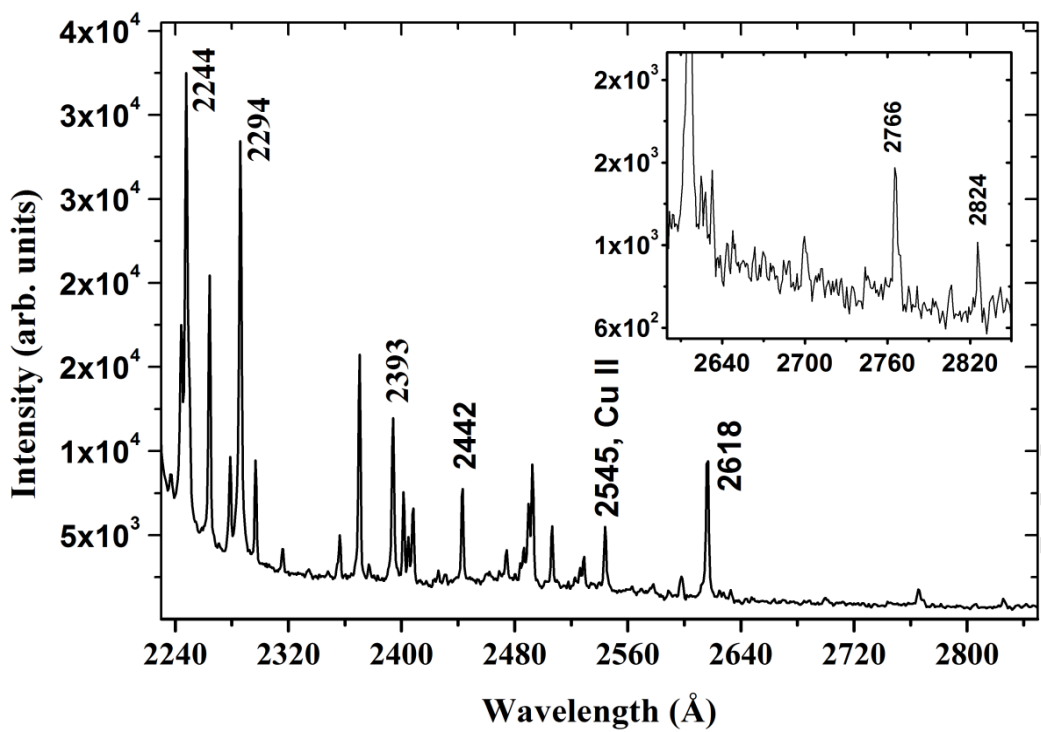

**Figure 8.** Emission spectrum of Cu arc in the 2230 – 2850 Å region recorded on a medium quartz spectrograph using IP.

## 4. Discussion

In the VUV region, the photostimulable phosphor or IP detector is a good option where position sensitive detection is desired. This method is fast and a large wavelength region (in the present case ~ 250 Å) can be covered simultaneously. This is particularly useful in studies where possible drifts in experimental or beam conditions make it desirable to have a fast recording of the spectrum. For instance, the pressure dependent study on $N_2O$ (section 3.4) which would take an entire day with the PMT detection system can be performed in ~ 3 hours with the IP detector using the single plate holder and ~30 min using the multi-plate holder. In general, any spectra can be recorded even at very low beam currents (< 5mA), unlike in the case of PMT detection where one has to work with beam currents of > 50 mA for a good S/N ratio. The response of the IP is found to be good over a large range of synchrotron beam current ranging from 5 mA to 90 mA. We may mention here that it was found preferable to work at relatively lower beam currents, as the IP gets saturated at high beam currents even for a very small exposure time (< 1sec). Detection of VUV light by PMTs typically requires a sodium salicylate scintillator for conversion of VUV to visible radiation, while IPs are inherently sensitive in this region. Another advantage of the IP over the PMT arises due to the fact that it is not sensitive in visible regions. Thus background level mainly composed of scattered/stray light having a large visible component is reduced greatly.

The problems of electronic noise and pick-up present in detectors like PMTs, CCDs, MCPs, etc. are absent in the IP system as no electronics is involved in the detection process. IPs are also lower in cost compared to CCDs and MCPs. In comparison to the traditional position sensitive detection technique of photographic plates, the IP is much more convenient to use as it does not require any wet chemical processing or dark room facility. Moreover the lack of availability of photographic plates and the re-usability of the IP make it an attractive alternative. It may be noted that IPs are flexible, thus rendering it possible to place them along the focal plane of the spectrometer. Properties of PMTs and some of the commonly used position sensitive detectors in the VUV region are compared in Table I. It may be noted that for experiments which demand better resolution, one can switch over to the PMT detection which is conveniently interchangeable in the HRVUV beamline. One of the limitations of currently available IPs is that the resolution achievable is limited by size of the laser tip used for scanning which is 42 μm in the present case. However this number may improve with technological advances in future. Another point is that the VUV optimized plates require very careful handling and storage as they do not have any protective over-coating. Any scratches/fingerprints on the surface or over exposure to light or humidity can reduce the sensitivity and re-usability of the plates.

**Table I. Comparison of various detectors used in the vacuum ultraviolet region**

| Parameter | PMT | Photographic plate[a] | MCP[a] | IP[a] |
|---|---|---|---|---|
| Response | Linear | Nonlinear | Linear | Linear |
| Dynamic range | 4 orders | 2 orders | 2-3 orders | 4-5 orders |
| Spatial resolution | ~1 μm | ~ 50–100 μm | ~ 25–50 μm | ~ 50 μm |
| Durability | Rugged | Fragile | Fragile | Rugged, but careful handing required for TR plates |
| Spectral coverage | - | 250–300 mm | 20–50 mm | 200–400 mm |
| Usage | reusable | one time use | reusable | reusable |
| Electronic Noise | present | absent | present | absent |
| Recording time[b] | ~30–120 min | ~ 1–60 s | ~ 1–60 s | ~ 1–60 s |

[a]Ref [6]; [b]For a typical scan of ~200 Å in a 6.65 meter spectrometer

## 5. Conclusions

An image plate based detection system for detection of VUV radiation has been installed on the HRVUV beamline at the synchrotron radiation source Indus-1. The performance of the IP in terms of sensitivity, wavelength coverage, reproducibility and resolution have been demonstrated by recording photoabsorption spectra of Xe, $O_2$, $SO_2$ and $N_2O$ in the region 1150–2300 Å. The emission spectrum of Cu arc recorded in the 2200–3000 Å region shows a fairly good sensitivity up to ~ 2650 Å. Compared to the PMT detection method, the IP is found to be much more sensitive to low light levels, i.e. low beam currents and the time taken to record spectra also has been drastically reduced. Incorporation of the IP detector is expected to enhance the utilization of the HRVUV beamline particularly for experiments on transient species where the number densities are low and experimental conditions are subject to drifts. Although earlier researchers have demonstrated the capabilities of these plates in the VUV region, to the best of our knowledge usage of IPs at VUV synchrotron beamline facilities has not been reported. In the present study we have demonstrated the capabilities of the IP as an efficient detector for synchrotron radiation based VUV photoabsorption studies.

**Acknowledgements:**

The authors thank the Board of Research in Nuclear Sciences (BRNS) and Department of Atomic Energy for funding this project under the DAE-BRNS scheme. We gratefully acknowledge the role of Dr. B.N. Jagatap in initiating this collaborative project and also the valuable support provided by Dr. N.K. Sahoo and Dr. S.N. Jha during the course of this work. We thank Dr. R. D'Souza for useful suggestions and the A&MPD workshop staff for help rendered in mechanical fabrication jobs.